\documentclass{optica-article}

\journal{opticajournal} % for journals or Optica Open

\articletype{Research Article}

\usepackage{lineno}
\usepackage{siunitx}
\usepackage{amsmath}
\usepackage{mathrsfs}
\usepackage{mathtools}
\usepackage[T1]{fontenc}
\usepackage{graphicx}% Include figure files
\usepackage{dcolumn}% Align table columns on decimal point
\usepackage{bm}

\global\long\def\infint#1{\intop_{-\infty}^{+\infty}\:\mathrm{d}#1}%

\newcommand{\QOT}{Centre for Quantum Optical Technologies, Centre of New Technologies, University of Warsaw, Banacha 2c, 02-097 Warsaw, Poland}

\begin{document}

\title{Ultrafast electro-optic Time-Frequency Fractional Fourier Imaging at the Single-Photon Level}

\author{Michał Lipka,\authormark{1,*} and Michał Parniak\authormark{1}}

\address{\authormark{1}\QOT\\
}

\email{\authormark{*}mj.lipka@uw.edu.pl} %% email address is required; see note below about the corresponding author designation

% use {asbstract*} to suppress the copyright line. Copyright information will be added in production

\begin{abstract*} 
    The Fractional Fourier Transform (FRT) corresponds to an arbitrary-angle rotation in the phase space, e.g. the time-frequency (TF) space, and generalizes the fundamentally important Fourier Transform. FRT applications range from classical signal processing (e.g. time-correlated noise optimal filtering) to emerging quantum technologies (e.g. super-resolution TF imaging) which rely on or benefit from coherent low-noise TF operations. Here a versatile low-noise single-photon-compatible implementation of the FRT is presented. Optical TF FRT can be synthesized as a series of a spectral disperser, a time-lens, and another spectral disperser. Relying on the state-of-the-art electro-optic modulators (EOM) for the time-lens, our method avoids added noise inherent to the alternatives based on non-linear interactions (such as wave-mixing, cross-phase modulation, or parametric processes). Precise control of the EOM-driving radio-frequency signal enables fast all-electronic control of the FRT angle. In the experiment, we demonstrate FRT angles of up to \SI{1.63}{\radian} for pairs of coherent temporally separated \SI{11.5}{\pico\second}-wide pulses in the near-infrared (\SI{800}{\nano\meter}). We observe a good agreement between the simulated and measured output spectra in the bright-light and single-photon-level regimes, and for a range of pulse separations (\SI{20}{\pico\second} to \SI{26.67}{\pico\second}). Furthermore, a tradeoff is established between the maximal FRT angle and bandwidth, with the current setup accommodating up to \SI{248}{\giga\hertz} of bandwidth.
    With the ongoing progress in EOM on-chip integration, we envisage excellent scalability and vast applications in all-optical TF processing both in the classical and quantum regimes. 
\end{abstract*}

%%%%%%%%%%%%%%%%%%%%%%%%%%  body  %%%%%%%%%%%%%%%%%%%%%%%%%

\section{\label{sec:into}Introduction}

The time-frequency (TF) domain plays a fundamental role from classical photonic communication, extensively relying on technologies like wavelength-division-multiplexing, to emerging quantum technologies \cite{Li2023spectrally,Zhang2021,Albarelli2022,He2022,Lipka2021prl,Lipka:21,Yang2018,Lu:22}, for which complete TF frameworks have been proposed \cite{Brecht2015}. The latter in particular include super-resolution imaging \cite{Donohue2018, Shah2021}, mode sorting \cite{Joshi2022}, continuous-variables protocols \cite{Fabre2020,Fabre2022}, bandwidth shaping \cite{Karpiński2017,Sosnicki2023}, spectroscopy \cite{Lipka2021prl}, multi-mode quantum repeaters \cite{Yang2018}, and TF imaging systems serving as waveform compressors \cite{Foster2009} or optical oscilloscopes \cite{Foster2008}, also in one-shot configurations \cite{Hernandez2013}. 
The need for diverse ultrafast TF characterization methods grows with the increasing prevalence of femtosecond-laser-based techniques in biology, chemistry, and spectroscopy \cite{Kukura2007}, communications and medicine \cite{fermann2002ultrafast}, material science \cite{Sugioka2014}, quantum information and atomic and molecular physics \cite{deVivie-Riedle2007}. 
Among these, TF protocols in the quantum regime most often require precisely controlled time-frequency transformations that are sufficiently low-noise for the single-photon-level light.

The Fractional Fourier Transform (FRT) corresponding to an arbitrary rotation in the TF space constitutes an important operation vastly prevalent in classical signal processing e.g. for optimal filtering of time-correlated noise \cite{Kutay1997}, encryption \cite{Hennelly2003} or chirp-based encoding schemes \cite{Ouyang2016}. All-optical FRT processing avoids the digitization of an optical signal, often detrimental in terms of the system bandwidth and noise figures. Furthermore, electro-optic FRT has been theoretically proposed as a pre-distortion method mitigating temporal pulse broadening \cite{Han2011} and as a central element of secure chaos-based communication \cite{Cheng2014}.

Here we experimentally present a scalable, all-electronically-controlled implementation of a low-noise single-photon-compatible FRT based on a grating pulse stretcher followed by an electro-optic modulator (EOM) acting as a temporal lens. The implementation is aimed at the high-bandwidth regime of faint single-photon-level picosecond pulses where an ability to coherently transfer the elusive temporal structure to the much more easily accessible spectral degree of freedom is greatly valued. 

For optical TF processing, the FRT constitutes a fundamental, primitive operation - a coherent rotation of the TF coordinates. As such its applications are wide. 
In the quantum regime, one example is the TF implementations of super-resolution protocols based on mode-sorting. Indeed a constellation of $N$ parallel FRT operations with the FRT angles separated by $2\pi/N$ and followed by a passive linear optical network can implement an $N$ mode sorter \cite{Zhou2017}.
In the context of ultrafast pulse characterization, FRT combined with a projective spectral measurement provides a tomographic cross-section of the TF space, while combined with spectral interferometry, it allows for the retrieval of otherwise elusive cubic spectral phase coefficient \cite{Brunel2007}. Porting the ideas from conventional signal processing techniques also holds a promise to provide new applications, in particular related to filtration of time-correlated noise directly at the level of an optical signal. Such processing directly in the optical domain is particuarly important in the photon-starved regime.

Optical FRT has been previously demonstrated with a single dispersive element \textendash{} a linearly chirped Bragg diffraction grating (LCFBG) implementing a TF analog of the Fresnel diffraction \cite{CuadradoLaborde2013}. However, this conceptually and practically simple implementation remains constrained by poor frequency resolution, limited bandwidth, and lack of further tunability once the LCFBG is manufactured. An interesting alternative has been demonstrated in ref. \cite{Schnebelin2017} which describes a feedback-loop-based single-element FRT for optically-carried RF waveforms. The bandwidth is however limited by the loop response time to tens of MHz. A controllable TF FRT has been also implemented in a quantum memory, aimed at sub-GHz-bandwidth light \cite{niewelt2023}. For ultrafast high-bandwidth pulses an adjustable, however, technically demanding implementation of TF FRT includes a chain of a dispersive element (pulse stretcher), quadratic temporal phase modulation (a time lens), and another dispersive element \cite{Lohmann1994}. 
Hitherto demonstrations of ultrafast TF FRT relied on non-linear phenomena such as wave-mixing, cross-phase modulation, or parametric processes to implement the temporal lens, achieving large modulation depth, however at the cost of introducing inherent optical noise detrimental to the single-photon-level operation  \cite{Salem2008,Zhou2022}.

The EOM-based implementation remains low-noise and single-photon-compatible at the same time offering a robust and versatile control of the FRT angle via all-electronic adjustments of the driving signal.
While an ordinary Fourier transform \textendash{} a special case of FRT has been demonstrated with an EOM-based time lens and a grating-based dispersive line \cite{Kauffman1994}, it has not been extended experimentally nor in the interpretation to the regime of arbitrary FRT angles. 
The task of which requires precise control over the EOM driving waveform. 
Finally, EO TF FRT promises excellent scalability and compactness if the diffraction-grating-based dispersive lines are exchanged for fiber-based Bragg diffraction gratings and the EOMs and linear optical elements are implemented on-chip, which is in the range of current technology.

The article is organized as follows: in section \ref{sec:frt} we introduce the Fractional Fourier Transform and its action in the time-frequency phase-space, also for a special case of an initial state of two temporarily separated pulses which is further studied experimentally. Section \ref{sec:exp} presents the experimental and simulation results alongside an overview of the setup. Details of the experiment, simulation, and calculations are presented in section \ref{sec:methods}. In particular, we describe therein the methodology of the design and calibration of the optical and RF setups and discuss the bounds on the FRT bandwidth. Finally, we draw conclusions in section \ref{sec:conc}.

\begin{figure}[ht!]
\centering\includegraphics[width=1\columnwidth]{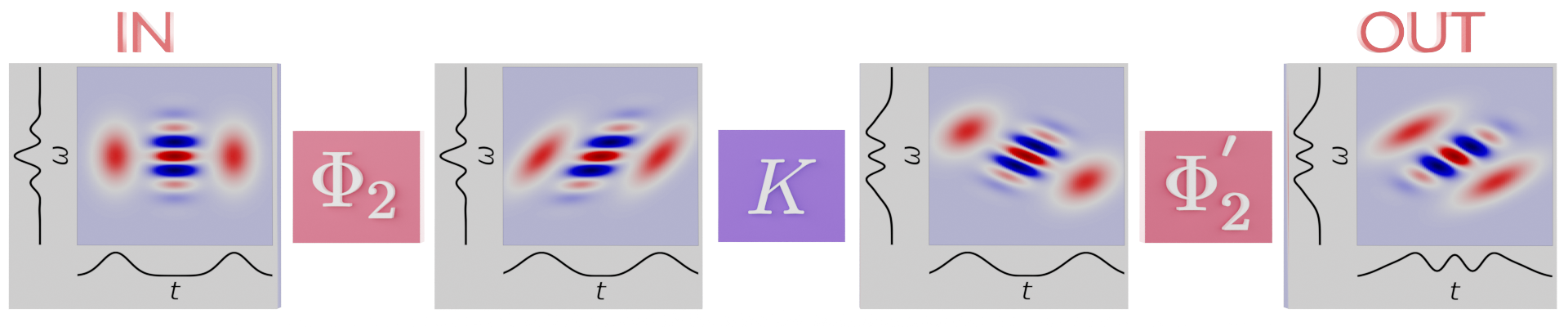}
\caption{Schematic depiction of a Fractional Fourier transform in the optical domain. A subsequent application of a quadratic phase in spectral $\Phi_2$, temporal $K$, and spectral $\Phi^{'}_2$ domains yields an arbitrary rotation of the initial state in the time-frequency space.
}
\label{fig:idea}
\end{figure}

\section{\label{sec:frt}Fractional Fourier Transform}
Fractional Fourier Transform (FRT) is a natural extension of a Fourier transform \textendash{} a transformation fundamental for (optical) signal processing \cite{Namias1980, Lohmann1993, Ozaktas1995, Sejdic2011}. 
The definition of FRT as an integral transform (see e.g. ref. \cite{Lohmann1993}) is not directly interpretable in terms of physical operations and as such we shall not repeat it here. 

Instead, a more insightful way to define FRT is as a propagator of the quantum harmonic oscillator. The behavior of which can be simply described as a rotation of the state's quasi-probability distribution in its TF phase space. In this picture, FRT of degree $P$ acts as a rotation by an angle $\alpha=P\pi/2$ \cite{Lohmann1993} with the special case of $P=1$ being the ordinary Fourier transform. To this end, let us describe the TF phase space with the Chronocyclic Wigner Function (CWF) defined for a well-behaved function $f(t)$ as \cite{Paye1992}:
\begin{equation}
    \mathcal{W}_f(\omega,t) =
    \infint t'\ f\left(t + \frac{t'}{2}\right)f^{*}\left(t - \frac{t'}{2}\right)e^{i\omega t'}.
\end{equation}

Conveniently, this rotation can be accomplished using alternating temporal and spectral shearing operations, with two possible sequences: temporal-spectral-temporal or spectral-temporal-spectral operations \cite{Lohmann1993}.
We are concerned with the latter case which at the level of light's electric field corresponds to imposing a quadratic phase in spectral 
\begin{equation}
\varphi_\omega(\omega)=\Phi_2(\omega-\omega_c)^2/2,
\label{eq:phiomega}
\end{equation}
temporal 
\begin{equation}
\varphi_t(t)=K t^2/2,
\label{eq:phit}
\end{equation}
and again spectral domains ($G^{'}_\omega$, $\omega^{'}_c$), where $\omega_c$, $\omega^{'}_c$ are some central frequencies, the time frame is aligned to the temporal centroid of the pulse, $\Phi_2$ is the group delay dispersion (GDD), and $K$ is the time-lens chirp rate. The sequence and the CWF at different stages have been schematically depicted in Fig.~\ref{fig:idea} for an initial state of two coherent, temporarily separated Gaussian pulses.  
Notably, these operations are completely analogous to a series of a propagation, passing through a lens, and a propagation of a single transverse spatial dimension in the paraxial approximation \cite{ Salem2013, Torres2011, Kolner1994}.

In our notation, the FRT angle $\alpha$ is given by the following equations \cite{Lohmann1993}

\begin{subequations}
\begin{align}
    & \Phi_2 = \mathcal{G} \tan{\frac{\alpha}{2}},
    \label{eq:alphdefeqsGom}\\
    & K = \mathcal{G}^{-1} \sin{\alpha},
    \label{eq:alphdefeqsGt}
\end{align}
\end{subequations}

where $\mathcal{G}$ is a GDD scaling factor. Solving for $\alpha$ we get
\begin{equation}
%\alpha =\pm 2\arctan\left(\frac{\sqrt{\Phi_2 K}}{\sqrt{2-\Phi_2 K}}\right),
\alpha =\pm 2\arcsin\sqrt{\frac{\Phi_2 K}{2}},
    \label{eq:frtangle}
\end{equation}
with the $+$ ($-$) sign for $0\leq\alpha<\pi$  ($-\pi<\alpha\leq 0$).
Notice that Eqs.~(\ref{eq:alphdefeqsGom}),(\ref{eq:alphdefeqsGt}) hold if $K$ and $\Phi_2$ have the same sign, hence experimentally a positive dispersion must be matched with a positive time lens and vice versa.

Let us write the electric field of light pulse in the slowly varying envelope (SVE) approximation \cite{Agrawal2013}.
\begin{equation}
    \mathcal{E}(t) = \mathcal{A}(t)\exp(i\omega_0 t) =
   \frac{\exp(i\omega_0 t)}{\sqrt{2\pi}}\int \mathrm{d}\tilde{\omega} \:\tilde{\mathcal{A}}(\tilde{\omega})\exp{(i\tilde{\omega} t)},
    \label{eq:et_svea}
\end{equation}
where $\tilde{\omega}=\omega-\omega_0$ and $\omega_0$ is the central optical frequency. We can further transform just the SVE $\mathcal{A}(t)$ while bearing eq. (\ref{eq:et_svea}) in mind. 

We begin by imposing the spectral phase, as given by Eq. (\ref{eq:phiomega}). 
%\begin{equation}
$\tilde{\mathcal{A}}(\tilde{\omega})\rightarrow \tilde{\mathcal{A}}(\tilde{\omega})\times\exp\left[{i\varphi_\omega(\omega)}\right]$.
%\end{equation}
Observe that any misalignment $\omega_c-\omega_0\neq 0$ between the central optical frequency of the pulse $\omega_0$ and the central frequency of the first stretcher $\omega_c$ effectively amounts to an additional linear spectral phase
\begin{equation}
\frac{\Phi_2}{2}(\omega-\omega_c)^2 = \frac{\Phi_2}{2}\left[\tilde{\omega}^2 - 2\omega(\omega_c-\omega_0)\right] + \mathrm{const.}
\end{equation}
which in the time domain manifests as an additional delay. In practice, such a misalignment will be compensated for when the delay of the EOM RF driving signal is matched to the arrival time of the pulse. Under such an assumption, the SVE after the stretcher reads:
\begin{equation}
    \mathcal{A}_1(t)=\frac{1}{\sqrt{2\pi}}\int \mathrm{d}\tilde{\omega} \:
    \tilde{\mathcal{A}}(\tilde{\omega})
    \exp{\left(i \frac{\Phi_2}{2}\tilde{\omega}^2\right)}
    \exp{\left(i\tilde{\omega} t\right)}.
\end{equation}
Finally, EOM modulation imposes a temporal phase according to Eq. (\ref{eq:phit}) $\mathcal{A}_2(t)=\mathcal{A}_1(t)\times\exp{\left[i\varphi_t(t)\right]}$.
In the experiment, we will observe the pulses with a spectrometer, which up to its limited resolution measures: 
\begin{subequations}
    
\begin{align}
    I_\mathrm{out}(\omega)=&|\tilde{\mathcal{A}}_2(\tilde{\omega})|^2,\\
    \tilde{\mathcal{A}}_2(\tilde{\omega})=&\frac{1}{\sqrt{2\pi}} \int \mathrm{d} t \: \mathcal{A}_1(t)
    \exp{\left[i\varphi_t(t)\right]}
    \exp{\left(-i\tilde{\omega} t\right)}.
    \label{eq:a2om}
\end{align}
\end{subequations}
Hence, we can neglect the last stretcher which ideally introduces only a spectral phase component, not observable in the spectrally-resolved intensity measurement.

Note that we have included only the quadratic terms in the spectral and temporal phase modulations. Higher-order terms become particularly important with higher bandwidth (spectral) or longer pulses (temporal). We further discuss the regime in which our approximation remains valid. For a rigorous description of the higher-order terms in a similar task of spectral-temporal imaging see ref. \cite{Bennett2001}.

\subsection{Two pulses}
Let us consider a special case of two identical pulses separated in time
\begin{equation}
\mathcal{A}(t) = \frac{1}{\sqrt{2}}\left[a(t-\frac{\delta t}{2})+e^{i\varphi}a(t+\frac{\delta t}{2})\right],
\end{equation}
where $a(t)$ describes the SVE of a single pulse, $\varphi$ their phase difference, and $\delta t$ the temporal separation. 
Such an input state, analogous to a cat state in quantum optics, constitutes a simple and traceable yet informative probe of the FRT setup parameters. 
Indeed the spectrum of the input state consists of a series of fringes, their density dependent on $\delta t$, while the spectra of the constituent pulses are identical. 
\begin{equation}
    |\tilde{\mathcal{A}}(\tilde{\omega})|^2=|\tilde{a}(\tilde{\omega})|^2\left[1+\cos{\left(\tilde{\omega} \delta t  + \varphi \right)}\right]
    \label{eq:initspectrum}
\end{equation}
Rotation in the TF space spectrally separates both pulses, gradually removing the fringes.
Let us assume Gaussian pulses with spectral width $\sigma$:
\begin{equation}
    a(t)=\left(\frac{\sigma^2}{\pi}\right)^{1/4}\exp{\left( -\frac{1}{2}t^2\sigma^2\right)}.
    %\tilde{a}(\tilde{\omega})=\frac{1}{(\pi\sigma^2)^{\frac{1}{4}}}\exp{\left( -\frac{\tilde{\omega}^2}{2\sigma^2}\right)}.
\end{equation}
Evaluating the integrals of Eq. (\ref{eq:a2om}) we get the following spectral intensity:
%\begin{subequations}
\begin{equation}
    \frac{1}{\sqrt{\pi \xi \sigma^2 }} \times
    \exp{\left(-\frac{\tilde{\omega} ^2}{\xi  \sigma ^2}-\frac{\delta t^2 K^2}{4 \xi  \sigma ^2}\right)} 
    \times  
    \left\lbrace\cosh
   \left(\frac{\delta t \tilde{\omega}  K}{\xi  \sigma ^2}\right)+\cos \left[\frac{\delta t \tilde{\omega} 
   }{\xi }\left(1-K \Phi_2\right)+\varphi \right]\right\rbrace,
\end{equation}

where:
\begin{equation}
    \xi =\frac{K^2}{\sigma ^4}+\left(K \Phi_2-1\right){}^2.
\end{equation}
%\end{subequations}
Indeed, putting $K=0$ ($\xi=1$) correctly retrieves the initial spectrum, as given by Eq. (\ref{eq:initspectrum}).
Whereas for the FRT angle of $\alpha=\pi/2$ we get the ordinary Fourier transform $t \rightarrow \mathcal{G}\omega$. In this case, using Eqs. (\ref{eq:alphdefeqsGt}), (\ref{eq:alphdefeqsGom}), we have $K=\mathcal{G}^{-1}$, $\Phi_2=\mathcal{G}$, $\xi=1/(\sigma^4 \mathcal{G}^2)$ and the transformed spectrum reads:
\begin{equation}
\frac{\mathcal{G} \sigma}{\sqrt{\pi }}
\exp{\left(-\frac{\sigma^2 \delta t^2}{4}\right)} 
\exp{\left(-\sigma ^2  \tilde{\omega} ^2 \mathcal{G}^2\right)}\times 
\left[\cosh \left(\delta t
   \tilde{\omega}  \mathcal{G} \sigma ^2\right)+\cos (\varphi )\right],
\end{equation}
which contains (after some algebraic transformations) two Gaussians spectrally separated (in angular frequency) by $\delta t/\mathcal{G}$. Notably, there are no spectral fringes, as the dependence on $\omega$ under the cosine term vanishes.

\begin{figure*}[ht!]
\centering\includegraphics[width=1\columnwidth]{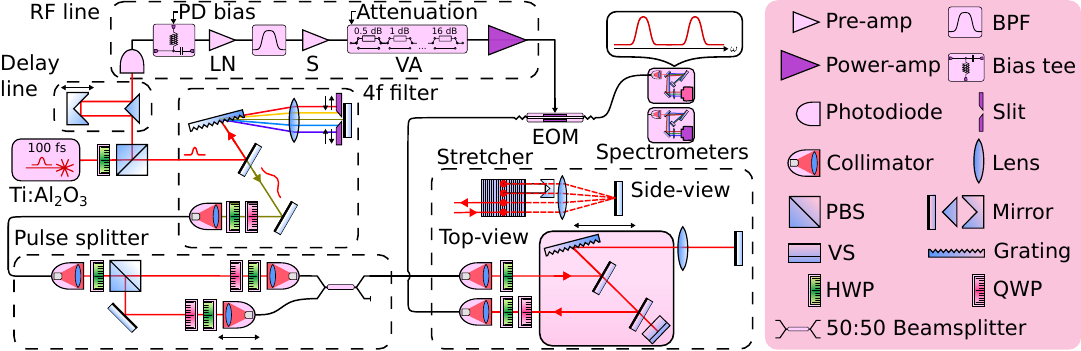}
\caption{Simplified experimental setup. Ultrafast ca. \SI{100}{\femto\second} pulses from a Ti:Al$_2$O$_3$ laser are spectrally filtered to ca. \SI{74}{\giga\hertz} (4f filter). The pulses are split and recombined in a Mach-Zehnder-type interferometer with a controlled arm-length difference (Pulse Splitter). The quadratic spectral phase is imposed by a grating-based quadruple-pass stretcher while the quadratic temporal phase by an electro-optic modulator (EOM). The EOM-driving signal is first obtained by exciting a photodiode with a small part (ca. average \SI{1}{\milli\watt}, \SI{80}{\mega\hertz} repetition) of the pre-filtered pulse. The radio-frequency (RF) line shapes and amplifies the signal. Spectra are measured either with a standard or a single-photon-level spectrometer. PBS \textendash{} polarizing beamsplitter, H(Q)WP \textendash{} true-zero-order half- (quarter-) waveplate, VS \textendash{} vertical-shift retroreflector, LN \textendash{} low noise, S \textendash{} standard, VA \textendash{} variable attenuator, BPF \textendash{} band-pass filter. 
}
\label{fig:setup}
\end{figure*}

\section{\label{sec:exp}Experiment}
\begin{figure*}[ht!]
\centering\includegraphics[width=1\columnwidth]{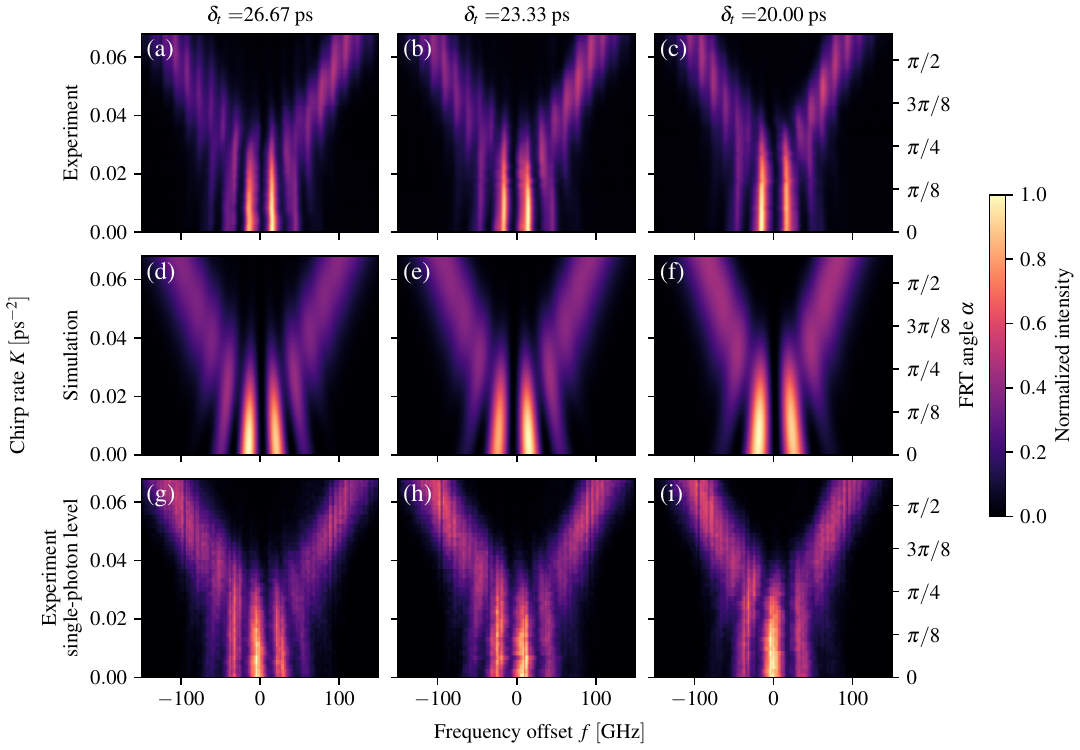}
\caption{
(a)-(c) Experimental spectra for a pair of temporally separated pulses undergoing FRT, for different levels of chirp rate $K$, with a fixed $\Phi_2=\SI{15.5}{\pico\second\squared}$. Equivalent FRT angles $\alpha$ are indicated. Each column corresponds to a different temporal separation of pulses $\delta t$.
(d)-(f) Corresponding simulated spectra. Fidelities $\mathscr{F}$ between experimental and simulated maps are (d) $96.28\%$, (e) $96.27\%$, (f) $95.88\%$. Simulation parameters were $f_m=\SI{15}{\giga\hertz}$, $\Delta_t=\SI{11.6}{\pico\second}$, $\Phi_2=\SI{15.5}{\pico\second\squared}$ (see section \ref{sec:methods}).
(g)-(i) Equivalent experimental spectra, collected at a single-photon level with an average number of photons per frame of $\bar{n}=0.38$ ($2.4\times10^{-4}$ photons per pulse) and $50\times10^3$ frames per each $K$ setting. Note a different phase $\varphi$, visible in the fringes.
\label{fig:results}
}
\end{figure*}

In the experiment, as depicted in Fig.~\ref{fig:setup}, we begin with \SI{100}{\femto\second} pulses from Ti:Sapphire laser (SpectraPhysics MaiTai) at a \SI{80}{\mega\hertz} repetition rate and a central wavelength of \SI{800}{\nano\meter}. With a 4f grating pulse shaper we apply an amplitude rectangular mask in the spectral domain to carve ca. \SI{74}{\giga\hertz} FWHM pulses which are coupled to a polarization-maintaining (PM) fiber and sent to a combined free-space and fiber Mach-Zehnder interferometer with a regulated delay in one arm. The interferometer splits the pulse into two temporarily separated parts with otherwise identical modes. One of the interferometer outputs is sent via fiber to the main FRT setup.
The FRT is implemented as a series of a grating stretcher (Martinez \cite{martinez1984negative} configuration with a single grating, quadruple-pass) and a fiber EOM. The output of the FRT is observed with a spectrometer. The efficiency of the FRT is ca. $1\%$ excluding detection.

While in principle the EOM would be followed by another stretcher, its quadratic spectral phase is not measurable with a spectrometer (see sec. \ref{sec:methods}).

%\subsection{Experimental spectra}

Experimentally collected spectra with bright light, for a range of FRT angles $\alpha$, have been depicted in Fig. \ref{fig:results}(a)-(c) together with a simulation (d)-(f), partially optimized for the highest fidelity $\mathscr{F}$. Equivalent single-photon-level measurements are depicted in Fig. \ref{fig:results} (g)-(i).

Note a very good agreement between the experimental spectra and the simulation. Residual infidelity stems mostly from an artificial spectrum modulation, visible in the strong-light (a)-(c) measurements, albeit not in the single-photon level measurements (g)-(i), and most probably stemming from a polarization-degree interference in the spectrometer.

Even with a relatively low $\Phi_2$ which can be easily extended several times in the current setup, the range of FRT angles $\alpha$ goes beyond the point of ordinary Fourier transform $\alpha=\pi/2$.

\section{\label{sec:methods}Methods}

\subsection{Simulation}
A direct numerical calculation has been performed to compare against experimental data. We assumed Gaussian, Fourier-limited pulses with intensity temporal FWHM of $\Delta_t=\SI{11.6}{\pico\second}$. The stretcher is assumed to produce only quadratic spectral phase, while for the EO modulator, we model the phase as a cosine with a main frequency, as experimentally measured, of $f_m=\SI{15}{\giga\hertz}$ and the amplitude equivalent of $K$ under a series expansion of cosine to the quadratic degree.
The simulation procedure consists of generating the SVE of the pulses on a temporal grid and a subsequent series of fast Fourier transforms intertwined with multiplications (in the matching domain) by spectral or temporal phase profiles corresponding to the action of a stretcher or a time lens, respectively.

\subsection{Fidelity} Fidelity $\mathscr{F}$ as a measure of correspondence between the simulated $\mathcal{I}^{(\mathrm{th})}(f,K)$ and experimental spectra $\mathcal{I}^{(\mathrm{exp})}(f,K)$ has been calculated as follows:
\begin{equation}
    \mathscr{F} = \frac{\sum_{f,K}\sqrt{\mathcal{I}^{(\mathrm{exp})}(f,K)\mathcal{I}^{(\mathrm{th})}(f,K)}}{\sqrt{\sum_{f,K}\mathcal{I}^{(\mathrm{exp})}(f,K)}\sqrt{\sum_{f,K}\mathcal{I}^{(\mathrm{exp})}(f,K)}},
\end{equation}
where the summation goes over the discrete points of the measured and simulated spectra in frequency $f$ and chirp rate $K$ coordinates.

The simulation parameters were taken as independently measured in the experiment ($\Phi_2$, range of $K$, $f_m$) or were optimized over for the highest fidelity (relative phase between pulses $\varphi$, the temporal mismatch between the time lens center and the pulses, and pulses FWHM $\Delta_t$).

\subsection{Second stretcher}
A complete FRT setup includes a second stretcher after the temporal lens to correct the spectral phase of the state. Its inclusion is vital with temporal or spectro-temporal characterization methods, as exemplified in the last two panels of Fig.~\ref{fig:idea}, and implemented for sub-GHz pulses in ref.~\cite{niewelt2023}. However, with a spectral intensity measurement, it does not affect the final results, except for the reduced efficiency of the setup. Hence we have experimentally verified the equivalence of the collected spectra with and without the last stretcher for a few cases and further proceeded without for the results presented herein. Finally, we note that simultaneous spectral and temporal phase characterization in the \SI{100}{\giga\hertz} regime remains technically challenging, in particular with confidence and resolution sufficient to characterize the FRT operation. 

\subsection{Grating stretcher design}
The stretcher is designed with a single grating, in a configuration akin presented in ref. \cite{Lai1994}. As illustrated in Fig. \ref{fig:setup}, a complex of a diffraction grating (Newport 33067FL01-290R, \SI{1800}{\ln\over\milli\meter}, \SI{26.7}{\degree} blaze angle) D-shaped input and output mirrors and a vertical-shift retro-reflector is placed on a translation stage. The stage movement direction, and input, diffracted, and output beams are kept parallel, allowing simple regulation of the GDD with the stage position. Input D-shape mirror and the grating angles can be adjusted together to set the frequency of the shortest optical path $\omega_c$ and the GDD scaling per unit length. 

A first-order diffracted (in the horizontal plane) beam travels over distance $L\approx\SI{71.5}{\milli\meter}$ and then enters a unit-magnification telescope with a mirror placed in the Fourier plane of the imaging lens (focal length $f$). Importantly, the lens is vertically shifted to spatially separate the first returning beam. Vertical-shift retro-reflector sends the returning beam back through the grating-telescope setup. The vertical positions are chosen, so that the last returning beam (4th pass) can exit via an output D-shape mirror. 

Following ref. \cite{WeinerUltrafastCh4} the total GDD of such a stretcher is given by:
\begin{equation}
    \Phi_2 = \frac{ m^2 \lambda^3 L}{2 \pi c^2 d^2 \cos^2\theta_d},
    \label{eq:gdd}
\end{equation}
where $m$ is the diffraction order, $c$ the speed of light, $d$ the grating groove period, and $\theta_d$ the diffraction angle for $\omega_c$. 

The choice of the focal length $f$ does not directly affect the GDD; however, the quadruple-pass design requires highly off-axis passage through the lens. Hence, care should be taken to avoid spherical aberration, especially present with lower-$f$ lenses. Ideally, an aspheric lens should be used. Additionally, if $L\gg f$ a precise calibration of the telescope becomes crucial. Simple ray optics consideration shows that the displacement of the mirror from the lens Fourier plane $\delta x$ must be kept $\delta x \ll f^2/L$.
With a narrow-band $\approx\SI{100}{\giga\hertz}$ light chromatic aberrations are less concerning. However, with wide-band pulses achromatic lenses are required, together with sufficiently large grating dimensions to accommodate the spread of the returning beam. 

Since with larger $f$ the optical path in a grating stretcher can get quite long (in our case with $f=\SI{200}{\milli\meter}$ it is ca. \SI{2}{\meter}) a larger beam with long Rayleigh range is generally preferred. 

Fundamentally GDD of a grating stretcher has a geometrical origin \cite{martinez1984negative}. Since only one spatial dimension acts as a proxy for the spectral degree of freedom, the output beam, even if correctly devoid of spatio-spectral correlations, will develop astigmatism. This is particularly prevalent for larger GDD $\gg\SI{100}{\pico\second\squared}$ and requires optical correction for efficient fiber coupling e.g. a cylindrical-lens-based telescope.

For the two implemented grating stretchers the efficiencies are 8\% (the one used in this work) and 13\%. Quadruple-pass through the grating itself amounts to ca. $(70\%)^4 = 24\%$ efficiency which is in agreement with the grating specification. The remaining losses are due to output fiber coupling. 

\begin{figure*}[ht!]
\centering\includegraphics[width=1\columnwidth]{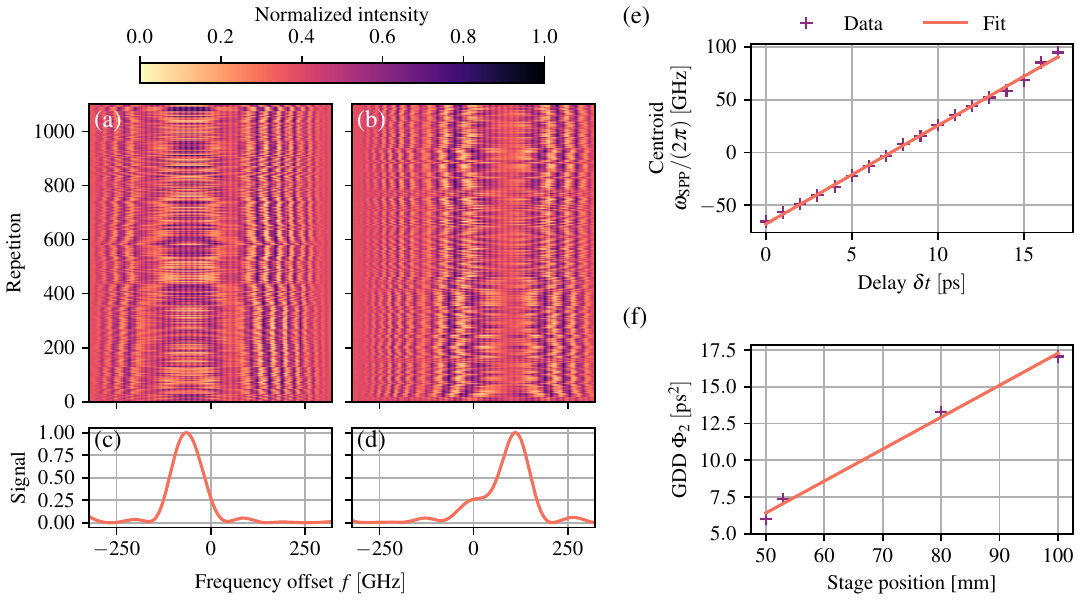}
\caption{
The stretcher GDD measurement via the stationary phase point method. (a)-(e) Data for a single position of the strecher stage (\SI{100}{\milli\meter}) and for a delay of either (a),(c) $\delta t=\SI{0}{\pico\second}$ or (b),(d) $\delta t = \SI{17}{\pico\second}$. (a),(b) Subsequent normalized interference spectra sampled at ca. \SI{100}{\hertz}. Frequency offset zero is arbitrary. (c),(d) Interference spectra are background-subtracted and Fourier-domain-filtered. Finally, normalized variance over the repetitions produces the illustrated signal which maximum corresponds to the stationary phase point. (e) Signal centroids (stationary phase points) for a series of delays. The slope of the linear fit corresponds to the measured GDD. (f) GDD measurement for a series of translation stage positions. Error bars are unnoticeable (typically 1\%-2\%).
}
\label{fig:gdd}
\end{figure*}
\subsection{Strecher GDD calibration}
Strecher GDD has been estimated via the stationary phase point (SPP) method \cite{sainz1994real} which is based on spectrally-resolved interferometry. The input pulse (spectrally wider than for the main FRT measurement) is divided into two parts with equal intensity. One passes through the stretcher while the other is temporarily shifted by a known small delay $\delta t$ (in addition to a major delay compensating the travel time through the stretcher). The two parts then interfere on a balanced beamsplitter which one output port is observed with a spectrometer.
The dominant spectral phase imposed by the stretcher is quadratic:
\begin{equation}
\varphi(\omega)=\frac{\Phi_2}{2}(\omega-\omega_c)^2,
\end{equation}
where $\omega_c$ is a frequency for which the optical path of the stretcher is the shortest. 
In turn, a delay of $\delta t$ corresponds to a linear spectral phase of $\delta t \omega$. Taking the phase difference between the arms and converting the quadratic (in $\omega$) polynomial to the canonical form, the observed fringe pattern will then have a form of    
\begin{equation}
    \propto \left(1+\cos\left\lbrace\frac{\Phi_2}{2}[\omega-(\omega_c+\frac{\delta t}{\Phi_2})]^2\right\rbrace\right),
\end{equation}
where we neglected the constant phase term and assumed perfect interference visibility. Hence, the stationary phase point can be observed as the point of vanishing fringe frequency $\omega_\mathrm{SPP}=\omega_c+\delta t/\Phi_2$. Estimating $\omega_\mathrm{SPP}(\delta t)$ for a series of $\delta t$ measurements allows $\Phi_2^{-1}$ to be retrieved as the linear fit coefficient. 
In our approach, for each $\delta t$ we collect many spectra with the global phase between the interferometer arms slowly fluctuating. This way, characteristic circular fringe patterns can be observed on a map of the spectrum versus repetition, as illustrated in Fig. \ref{fig:gdd} (a),(b). To estimate the SPP point, the maps are low-pass filtered along the frequency dimension (convolved with a Gaussian kernel) and a variance is calculated across repetitions. The maximum of the variance corresponds to the point where the fringe frequency vanishes (highest response to the filter) i.e. the SPP.         

Fig. \ref{fig:gdd} (e) depicts a final result for a series of $\delta t$. The slope of a linear fit corresponds to the estimated GDD $\Phi_2$. Fig. \ref{fig:gdd} (f) gives GDD estimates for different settings of the translation stage which effectively changes $L$ in eq. (\ref{eq:gdd}). GDD error taken as a standard deviation of the linear fit slope, seems to be in the range of 1\%-2\%; however, already a higher variance can be observed by slightly altering the signal filtering parameters. Hence, realistically we estimate the true GDD error in the range of 5-10\%, with higher errors for lower GDDs. In this regime, a more precise 2-dimensional SPP could in principle yield better estimates \cite{kovacs2005dispersion}.

\begin{figure*}[ht!]
\centering\includegraphics[width=1\columnwidth]{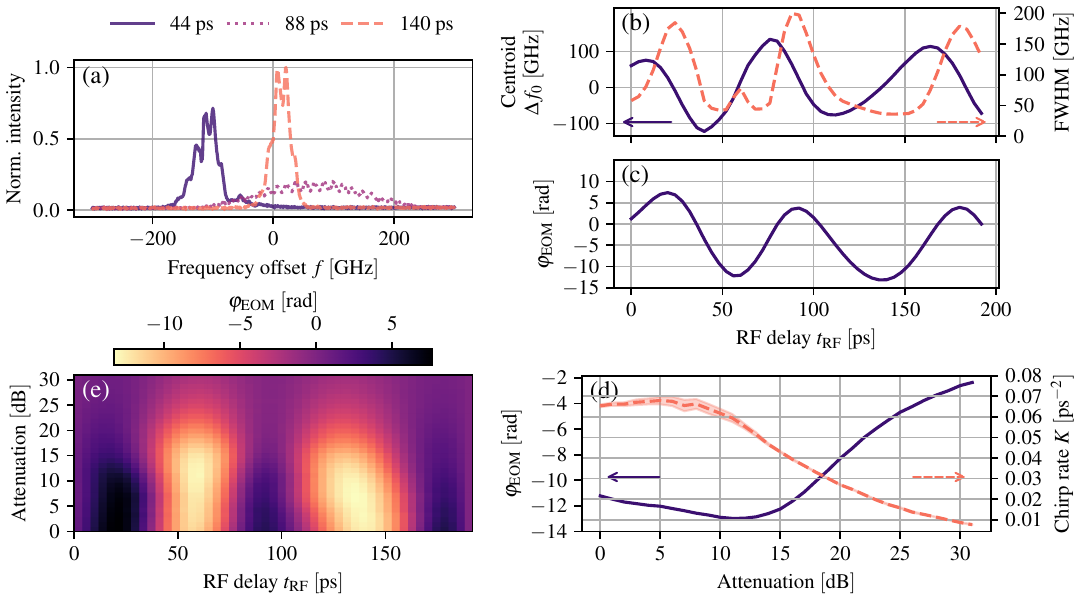}
\caption{
Optical measurement of the EOM driving waveform. (a) Spectra of a single pulse passing subsequently through the stretcher and the EOM for the attenuation of \SI{5}{\dB} and different RF delays $t_\mathrm{RF}$. (b) Centroid $\Delta f_0$ and full-width-half-maximum (FWHM) of Gaussians fitted to each collected spectrum. (c)-(e) Phase modulation of the EOM $\varphi_\mathrm{EOM}$ retrieved by integrating the centroid positions. (c) For a single attenuation of \SI{5}{\dB}. (d) (solid curve) For a single RF delay of $t_\mathrm{RF}=\SI{60}{\pico\second}$. (d) (dashed curve) chirp rate $K$ from a parabola fit to the waveform maximum for a given attenuation level. 
}
\label{fig:eom}
\end{figure*}
\subsection{RF line and EO modulation}
The radio-frequency (RF) signal driving the EO modulator (iXblue NIR-MPX800-LN-20) is produced with a photodiode (PD) (Hamamatsu Photonics G4176-03 + bias tee Mini-circuits ZX85-12G-S+ set to \SI{10}{\volt}) excited with the origin femtosecond pulses, preamplified (low noise Mini-circuits ZX60-06183LN+ and Mini-circuits  ZX60-183-S+), filtered ($6-18$\SI{}{\giga\hertz} band pass filter Mini-circuits ZBSS-12G-S+), amplitude-controlled with a programmable attenuator (6-bit, \SI{0.5}{\dB} LSB, Analog Devices ADRF5720), and finally, power-amplified with a \SI{3}{\watt} amplifier (Mini-circuits ZVE-3W-183+). The exact RF chain sequence is depicted in Fig. \ref{fig:setup}.

Delay matching between the optical pulse and the RF driving signal is twofold: via a motorized optical delay line for the PD driving pulse (Delay line) and via a ca. \SI{1}{\meter} long manual quadruple-pass delay line placed after the Pulse Splitter (not shown).

Conveniently, the RF driving pulse can be temporarily characterized by observing the spectral shift of a single pulse sent through the stretcher-EOM combination. The spectral shift $\Delta f_0$ is proportional to the temporal phase gradient
\begin{equation}
\partial_t\varphi_\mathrm{EOM}(t) = 2 \pi \Delta f_0.
\end{equation}
Hence, estimating the centroid  $\Delta f_0$ of measured spectra for a series of RF delays $t_\mathrm{RF}$ and integrating $\Delta f_0 (t_\mathrm{RF})$ over the delay $t_\mathrm{RF}$ we get the EOM-induced phase $\varphi_\mathrm{ROM}(t)$. To this end, such a measurement for the RF driving waveform has been depicted in Fig. \ref{fig:eom}.

The chirp rate $K$ can be directly retrieved by fitting a parabola to the waveform extrema. 
Otherwise, a theoretical prediction, obtained by series-expanding a cosine, reads:
\begin{equation}
    K = \frac{1}{2\pi^3}\frac{V_\pi}{f_m^2 V_\mathrm{pp}},
\end{equation}
where $V_\pi\;[\mathrm{V}]$ is the voltage on EOM for a $\pi$ phase shift, $V_\mathrm{pp}$ is the applied peak-to-peak voltage, and $f_m$ is the modulation frequency
In our case $V_\pi=\SI{4}{\volt}$, $V_\mathrm{pp}\approx\SI{19.5}{\volt}$, and $f_m\approx\SI{15}{\giga\hertz}$. Maximal $V_\mathrm{pp}\approx\SI{34.6}{\volt}$ limited by the \SI{3}{\watt} power amplifier, and RF power dissipation in the EOM.

Importantly, the ability to perform arbitrary-angle FRT without distortion relies on a precise RF amplitude control with a minimal introduced phased shift. In our implementation, it is warranted by a carefully selected variable attenuator. Even though its nominal relative phase shift is ca. \SI{20}{\degree} at \SI{15}{\giga\hertz} between minimal \SI{0}{\dB} and maximal \SI{31.5}{\dB} attenuation states (linearly increasing through intermediate states), we experimentally find the net phase shift much lower. It is visible in Fig. \ref{fig:eom} (e) where at $t_\mathrm{RF}\approx\SI{60}{\pico\second}$ the $\arg\max\varphi_\mathrm{EOM}$ stays constant regardless of attenuation. Conveniently, the attenuation level is all-digitally controlled with a Bluepill STM32F103C8T6 board directly selecting the attenuation state of the variable attenuator with a 6-bit range and a least-significant-bit (LSB) step of \SI{0.5}{\dB}.

\subsection {Pulse preparation and FRT bandwidth}

\begin{figure}[ht!]
\centering\includegraphics[width=0.7\columnwidth]{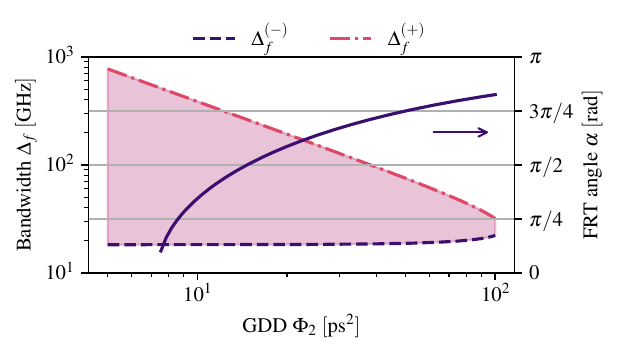}
\caption{
(left axis) Bounds (shaded region) on the initial pulse bandwidth $\Delta_f$ for a range of spectral GDD values $\Phi_2$, assuming a temporal aperture of $D_t=\SI{24.3}{\pico\second}$. (right axis) FRT angle $\alpha$ for a chirp rate of $K=\SI{6.80e-2}{\per\square\pico\second}$ and a range of $\Phi_2$.
}
\label{fig:bw}
\end{figure}

The bandwidth of EO FRT is mainly limited by the modulation frequency $f_m$ of the EOM which determines the temporal aperture of the time lens. In our case $f_m\approx\SI{15}{\giga\hertz}$.
%(with minor preamp replacements should reach \SI{18}{\giga\hertz}).
Assuming $5\%$ accuracy of the parabolic approximation, we get a full-width temporal aperture of ca. $\mathcal{D}_t=0.34\times f_m^{-1}\approx\SI{24.3}{\pico\second}$.
Noticeably $\mathcal{D}_t$ applies \textbf{after} the first stretcher, hence the pulse duration is limited after accommodating for elongation due to $\Phi_2$. Let us assume a Fourier-limited Gaussian pulse with temporal intensity full-width-half-maximum $\Delta_t$.
The elongated pulse will have a width of \cite{WeinerUltrafastCh4}
\begin{equation}
    \Delta^{\mathrm{post}}_t  = \Delta_t\sqrt{1+\left(4\log 2 \frac{\Phi_2}{\Delta^2_t}\right)^2}.
\end{equation}

We require $\Delta_f^\mathrm{post}\leq\mathcal{D}_t$. For Fourier-limited pulses $\Delta_t \Delta_\omega = 4 \log 2$, hence we get the following bounds for the initial pulse bandwidth in terms of the spectrum (intensity) full-width-half-maximum 
$\Delta_f=\Delta_\omega/(2\pi)$ 

\begin{subequations}

\begin{align}
      & \Delta_f^{(-)}  \leq \Delta_f \leq \Delta_f^{(+)},\\% &&\\
      & \Delta_f^{(\pm)} = \frac{\mathcal{D}_t}{2\pi\sqrt{2} \Phi_2} \sqrt{1\pm \sqrt{1- \left(\frac{8 \Phi_2 \log 2}{\mathcal{D}_t^2}\right)^2}}.%  && 
\end{align}

\end{subequations}
Note the limit of $\Delta_f^{(-)}$ for small $\Phi_2$:
\begin{equation}
    \lim_{\Phi_2\to0}\Delta_f^{(-)} = \frac{2\log 2}{ \pi \mathcal{D}_t},
\end{equation}
which just states that the initial pulse must fit in the temporal aperture. Similarly, for the range of bandwidths, we get asymptotically:
\begin{equation}
\Delta_f^{(+)}-\Delta_f^{(-)}\sim \frac{1}{2\pi}\frac{\mathcal{D}_t}{\Phi_2}
\end{equation}
Conversely, for large $\Phi_2$ such that 
$\xi\coloneqq 1- \left(8 \Phi_2 \log 2/\mathcal{D}_t^2\right)^2\xrightarrow{}0$
we get:
\begin{equation}
\Delta_f^{(+)}-\Delta_f^{(-)}\xrightarrow{}\frac{2 \sqrt{2}  \log (2)}{\pi }\times\frac{\sqrt{\xi}}{d}.
\end{equation}

The bounds on the initial pulse bandwidth have been depicted in Fig. \ref{fig:bw} together with the FRT angle, as given by Eq. (\ref{eq:frtangle}), for a fixed chirp rate $K=\SI{6.80e-2}{\per\square\pico\second}$.
For instance, taking 
$\Phi_2 = \SI{15.5}{\pico\second\squared}$, we get $\SI{18.2}{\giga\hertz}\leq\Delta_f\leq\SI{248.9}{\giga\hertz}$.
%$\Phi_2 = \SI{40}{\pico\second\squared}$, we get $\SI{18.5}{\giga\hertz}\leq\Delta_f\leq\SI{94.9}{\giga\hertz}$.
Lower $\Phi_2$ presents a wider bandwidth, yet for a given angle of FRT would require a higher $K$, which in turn for a fixed $f_m$ needs higher power on EOM or a smaller $V_\pi$.
Nevertheless, within the limitations of the current setup, $\Phi_2$ can be increased several times (just by increasing the relatively small distance $L$), trading off the maximal bandwidth $\Delta f$ for increased FRT angle $\alpha$. 

In the experiment, we use a folded 4f setup, illustrated in Fig. \ref{fig:setup}, to carve pulses with ca. $\Delta_f\approx\SI{74}{\giga\hertz}$. The setup involves a first-order diffraction on a grating, far-field imaged onto a rectangular 1-dimensional aperture with regulated width. Immediately after the aperture a mirror is placed facilitating backpropagation through the setup. The vertical offset of the imaging lens spatially separates the returning beam.

\subsection{Spectrometers}
In the strong-light regime for the spectrally-resolved intensity detection, we employ a custom-made grating spectrometer in a second-order double-pass configuration, described in detail in ref. \cite{Kurzyna2022}.

Whereas for the single-photon-level measurements, we use a similar grating spectrometer (\SI{1200}{\ln\per\milli\meter}, second-order, double-pass) with a custom single-photon camera, described in detail in refs \cite{Lipka:21, Lipka2021prl}, as the detector. A collimated beam with ca. \SI{10}{\milli\meter} diameter impinges onto the diffraction grating with the incidence angle $\theta_i$ of ca. \SI{67}{\degree} and a diffraction angle $\theta_d$ close to \SI{90}{\degree}. A vertical-shift retro-reflector routes the $m=2$ diffracted order back onto the grating, vertically-shifted by ca. \SI{10}{\milli\meter}. A diffracted beam from the second pass is separated via a D-shaped mirror, followed by a $f=\SI{400}{\milli\meter}$ lens and a single-photon camera placed in its focal range. The single-photon camera is based on a fast CMOS sensor (Luxima LUX2100) and custom FPGA electronics (Xilinx Zynq-7020, Z-turn board), with an image intensifier (II, Hamamatsu V7090D71G262) in the Chevron configuration, for single-photon sensitivity. The camera collects $50\times400$ pixel frames at $\SI{2e4}{}$ frames per second, with II gating time of $\SI{20}{\micro\second}$ . A camera pixel corresponds to a frequency step of \SI{1.67}{\giga\hertz}, calibrated independently with a series of interferometric fringe-density measurements for a pair of temporarily separated pulses with varying, known separation. 
The theoretical frequency resolution $\delta f$ of the single-photon spectrometer is limited by the diffraction grating and reads \cite{LoewenPopovDiffractionGratings}:
\begin{equation}
    \delta f = 2\times \frac{c}{W |\sin{\theta_i}+\sin{\theta_d}|} \leq \frac{c}{W},
\end{equation}
where $\theta_i$, $\theta_d$ are taken positive on the same side of the grating normal, $W\approx\SI{19.2}{\milli\meter}$ is the diameter of the grating area covered by the beam (including elongation due to the angle of incidence and diffraction), $c$ is the speed of light, and the factor of $2$ comes from a double-pass configuration. 
In our case $\delta f \approx \SI{8}{\giga\hertz}$.

The measured efficiency of the single-photon spectrometer was $\approx4\times10^{-5}$.

\section{\label{sec:conc}Conclusion}
We have demonstrated a scalable, single-photon-level-compatible method for performing an arbitrary-angle Fractional Fourier Transform in the Time-Frequency domain. Based on an electro-optic time lens, our FRT implementation avoids the optical noise inherent to the solutions based on non-linear processes, hence remains compatible with the single-photon-level quantum light. 
A tailored RF line with a carefully chosen variable attenuator provides an all-electronic, fast, and precise control over the FRT angle. Furthermore, this FRT implementation is based on optical devices with miniaturized on-chip or fiber-based equivalents, promising excellent scalability.

The single-photon-level compatibility, all-electronic control, and the potential for miniaturization make the FRT a suitable building block for TF-domain quantum optical protocols with prominent applications in mode-sorting, super-resolution imaging, or synthesis of arbitrary unitaries. 
Finally, it may find applications not only in the quantum domain, in the most advanced devices such as quantum time-frequency processor \cite{PhysRevLett.120.030502, PhysRevA.104.062437}, but also in engineering more complex and capable solutions for classical light control for the purpose of various mode-division multiplexing protocols \cite{Schroder:13}.

%%%%%%%%%%%%%%%%%%%%%%%%%%%%%%%%%%%%%%%%%%%%%%%%%%%%%%%%%% BACKMATTER
\begin{backmatter}
\bmsection{Funding}
Fundacja na rzecz Nauki Polskiej (MAB/2018/4 \textquotedblleft Quantum Optical Technologies\textquotedblright{}); European Regional Development Fund; Narodowe Centrum Nauki (2021/41/N/ST2/02926);
\bmsection{Acknowledgments}
The \textquotedblleft Quantum Optical Technologies\textquotedblright{} project is
carried out within the International Research Agendas programme of the
Foundation for Polish Science co-financed by the European Union under the
European Regional Development Fund.
This research was funded in whole  or in part by National Science Centre, Poland 2021/41/N/ST2/02926. 
ML was supported by the Foundation for Polish Science (FNP) via the START scholarship.
We would like to thank K. Banaszek and W. Wasilewski for the generous support.

\bmsection{Disclosures}
The authors declare no conflicts of interest.

\bmsection{Data availability} 
 Data for figures \ref{fig:results}, \ref{fig:gdd} and \ref{fig:eom}  has been deposited at \cite{our_data} (Harvard Dataverse).
 \end{backmatter}

%%%%%%%%%% If using BibTeX:
\bibliography{frt}

\end{document}